\documentclass[aps,letterpaper,amssymb,twocolumn,prl]{revtex4}

\usepackage{bm}
\usepackage{amssymb}
\usepackage{times}
\usepackage{latexsym}
\usepackage{graphicx}
\usepackage{color}
\usepackage{subfigure}
\begin{document}
\title{ Bright and Dark Solitons and Breathers in Strongly Repulsive Bose-Einstein Condensates } 
\author{ William P. Reinhardt$^{1,3}$, Indubala I. Satija$^{2,3}$, Bryce Robbins$^{4,3}$ and Charles W. Clark$^{3}$}
\affiliation{$^{1}$ Department of Chemistry, University of Washington, Seattle, WA 98195-1700}
 \affiliation{$^{2}$ Department of Physics, George Mason University, 
 Fairfax, VA 22030}
\affiliation{ $^{3}$ Joint Quantum Institute, National Institute of Standards and
Technology and University of Maryland, Gaithersburg, MD 20899}
\affiliation{ $^{4}$ Department of Physics, Colorado School of Mines, Golden, CO 80401}
\date{\today}
\begin{abstract}

Collisional dynamics of solitary matter waves of superfluid hard core bosons, consisting of 
dark and bright non-linear waves as well as
supersonic periodic trains, reveals  
remarkable richness and coherence, with the phase of the condensate playing a key role.
Depending upon the condensate density and their relative velocity, two distinctive collisional types
emerge: the intuitively expected repulsive collisions due to the hard core boson constraint; and, also
collisions in which they "pass through" one another.
In addition to confirming the soliton status of both bright and dark solitary waves, 
our studies reveal
a variety of multi-solitons including multiple families of breathers, 
that can be produced and precisely controlled via quantum phase engineering.
\end{abstract}
\maketitle
Solitary waves and solitons are encountered in systems as diverse as
classical water waves\cite{book}, magnetic materials\cite{NIST}, fiber-optic communication \cite{Agrawal}, as well as  
Bose-Einstein condensates (BEC)\cite{gpe}.
Rooted in nonlinearity which balances
dispersive effects, solitons are fascinating non-linear waves that encode coherence underlying the system.
Solitons are distinguished from solitary waves by their behavior in collisions.
Solitons remain intact after collision, which is usually associated with
the integrability of their non-linear dynamics.
Therefore, discovery of solitons in a complex
physical system suggests the existence of a hidden simplicity of the underlying nonlinear equations
of motion.       

Intrinsically nonlinear in nature due to inter-particle interactions, 
the BEC systems are natural fertile ground for exploring 
solitons. In dilute atomic gaseous BECs
which are simply described in terms of the properties
of the non-linear Schrodinger equation (NLSE), or Gross-Pitaevski equation (GPE), bright and dark solitons are
characterized not only by persistent density anomalies, but also by characteristic phase modulations 
across their profiles. The quantum phase playing quite different roles in the bright (attractive condensates)\cite{CBPRL} 
and dark (repulsive condensates), both of which have been experimentally realized by\cite{expt}, \cite{Bexpt}.

Here we report the existence of multiple species of solitons in strongly repulsive BECs as might arise in consideration of a
hard core boson (HCB) system on a 1D lattice \cite{PRL}. The presence of both dark (density notch) and bright (waves of elevation) solitons in the same physical system is a novelty in ultra-cold atomic systems as the well known solitons in weakly interacting gaseous BECs
are dark or bright depending upon the repulsive or the attractive nature of the interparticle interaction.\cite{gpe}
Considering collisions of like pairs (bright-bright or dark-dark) if their relative propagation speed exceeds a (density dependent) critical value, two colliding soliton adapt to the hard core constraint by broadening, and seemingly to pass through each other.
Below the critical speed, the solitons bounce of each other, and their density peaks (bright-bright) or minima (dark-dark) repel visibly,  mimicking the reflection of a wave from a wall, as the hard core density constraint stops them from passing through one another. In both types of collisions, the dynamics is controlled by the quantum phase
of the composite non-linear wave and the density is found to remain stationary during such a pair collision. 

When the soliton speed exceeds the speed of sound,
solitons occur as trains. When these supersonic trains collide, they always pass through each other.  

In contrast to the collisions involving members of the same soliton species ( dark-dark or bright-bright collisions), the
colliding solitons of different species ( dark-bright) always pass through each other. 
The special case near half-filling provides a collision scenario resembling the soliton-antisoliton {\it breather} pair of the sine-Gordon system.\cite{book} In this case, the two species are mirror images of each other, with respect to the ambient surface of constant density  $\rho_0$ with respect to which they move, and they disappear as perturbations in the density at the mid-point of the collision and re-emerge unscathed. Their exceptional stability is rooted in the
{\it conservation of quantum phase jump} through the collision. As we discuss below, such 
inter species collisions are characterized by a stationary phase profile, whereas the intra species collisions result in stationary density profile. As in the sine-Gordon case, the HCB system supports breathers: oscillatory bound states of dark and bright solitons.
In contrast to Sine-Gordon on the other hand, these breathers can be dissociated into dark and bright soliton pairs by tuning a system parameter, or as in the sine-Gordon case by small alterations in initial density or phase conditions.
We show that with appropriate quantum phase imprinting it is possible to create a families of breathers where the number, and oscillation frequency, of the breathers can be tuned by changing the phase profile.

The system under consideration here is the limiting case
of the extended Bose Hubbard model in $D$ dimensions,
\\
\begin{equation}
H=-\sum_{j,a}[t \,b_j^{\dagger} b_{j+a}+ V n_j n_{j+a}]\\+\sum_j U n_{j}
(n_{j}-1)
-(\mu-2tD)n_{j}
\label {BH}
\end{equation}
\\
Here,  $b_j^{\dagger}$ and $b_j$ are the creation and
annihilation operators
for a  boson at the lattice site $j$,  $n_{j}$ is the number operator, $a$
labels nearest-neighbor (nn) sites ,
$t$ is the nn hopping parameter,  $U$ is the on-site interaction strength,
and $\mu$ is the chemical potential.
An attractive nearest-neighbor interaction $V < 0$, is introduced to soften the effect of a strong repulsive onsite interaction,
$U >> 0$ and may also describe systems with long range interactions such as dipolar gases.\cite{Baiz}
In HCB limit ($U \rightarrow \infty$), no more than one boson can occupy a given site. 
In this case, we can represent the system by a lattice of spin-$1/2$ particles where two spin states
correspond to two allowed boson number state $|0>$ and $|1>$.
Thus, the HCB system is explicitly described by,
\begin{eqnarray}
H_S=-\sum_{j, a}[t\,\, {\bf{\hat{S}_j} \cdot \bf{\hat{S}_{j+a}}} -g\,\, \hat{S}_j^z \hat{S}_{j+a}^z]-
g\sum_j{\textstyle}\,\, \hat{S}_j^z
\label{mag}
\end{eqnarray}
where $g=t-V$ and the spin flip operators $\hat{S}^{\pm}=\hat{S}_x\pm i\hat{S}_y$ 
are the annihilation and the creation operators
of the corresponding bosonic Hamiltonian
,$b_j \rightarrow \hat{S}_{j}^{+}$. Thus the order parameter that describes BEC wave function is
$\psi_j^s=\langle S_j^+\rangle$. 
In this mean-field description, the evolution equation
for the order parameter is obtained by taking the spin-coherent state average of the
the Heisenberg  equation of motion.
The spin coherent state $|\tau_j>$ at each site $j$ can be parametrized as:\\
$|\tau_j>=e^{ i\frac{\phi_j}{2}} [e^{-i\frac{\phi_j}{2}} \cos \frac{\theta_j}{2} |\uparrow>+e^{i \frac{\phi_j}{2}} \sin \frac{\theta_j}{2}
 |\downarrow>$].
With this choice, the HCB system is mapped to a system of classical spins
where the particle density
$\rho_j$, and the condensate density $\rho_j^s$ 
satisfy $\rho^s_j= \rho_j\rho^h_j$, where 
$\rho^h_j = 1-\rho_j$ is the hole density. In this representation,
$\psi_j^s =
\sqrt{\rho_j^s} e^{i \phi}$. We cast the 
equations of motion in terms of the canonical variable $\phi$ and $\cos(\theta)=(1-2\rho)$, where for simplicity we
denote $\cos(\theta)$ equal to 
$\delta$ are:
\\
\begin{eqnarray}
\dot{\delta_j}&=&\frac{t}{2} \sum_a \sqrt{(1-\delta^2_j)(1-\delta^2_{j+a})} \sin (\phi_{j+a}-\phi_j)\\
\dot{\phi_j}&=&\frac{t}{2} \frac{\delta_j}{\sqrt{(1-\delta^2_j)}} \sum_a \sqrt{1-\delta^2_{j+a}}\cos(\phi_{j+a}-\phi_j)\\
&-&\frac{V}{2}\sum_a \delta_{j+a} - \frac{U_e}{2}\delta_0
\label{canonical}
\end{eqnarray}

In the continuum approximation, the equations have been shown to support solitary waves\cite{PRL} riding upon
a background density $\rho_0$: $\rho(z)=\rho_0+f(z)$,
with $z=x-vt$.
\begin{eqnarray*}
f(z, \rho_0)^{\pm}&=&\frac {2\gamma^2 \rho_0\rho_0^h}{\pm \sqrt{(\rho_0^h-\rho_0)^2+4\gamma^2 \rho_0 \rho_0^h}
\cosh \frac{z}{\Gamma}-(\rho_0^h-\rho_0)},
\label{feqn}
\end{eqnarray*}
Here $\gamma=\sqrt{1-\bar{v}^2}$ with $\bar{v}$ the speed of the solitary wave in units of the speed of sound velocity 
$c_s=\sqrt{2\rho_0^s(1-V/t)}$: $\Gamma$ is the width of the soliton,
$\Gamma^{-1}=\gamma \sqrt{{\frac{2(1-\frac{V}{t})\rho_0 \rho_0^h}{\frac{1}{4}(\rho_0^h-\rho_0)^2+\frac{V}{t}\rho_0 \rho_0^h}}}$.
The characteristic phase jump associated with the solitary waves is,
$\Delta \phi^{\pm} =\sqrt{1-2c_s^2} [\pm \sin^{-1} \frac{ 2\gamma \bar{v} (1-2\rho_0)}{1-4\rho^s_0\bar{v}^2}+\pi ]$.

The presence of two species of solitary waves is a direct consequence of particle-hole symmetry underlying the equations of motion.
The existence of $f(z,\rho_0)$ superposed on the background particle density $\rho_0$  implies the existence of a counterpart
$f(x, \rho_0^h)$, superposed upon a corresponding hole density $\rho_0^h$. In fact it is easy to see that
$f^{\pm}(z,\rho_0) = \pm f^{\mp}(z,\rho_0^h)$. For $\rho_0 < 1/2$, the $\pm$ corresponds to bright and dark
solitons respectively. The bright solitons have the unusual property of persisting at speeds upto the speed of sound, sharp contrast
to the dark species that resemble the dark soliton of GPE whose amplitude goes to zero at sound velocity.
Furthermore, unlike dark solitons, bright waves
, in their  density profile always include $\rho=1/2$, the branch point of the equation $\rho_s=\rho (1-\rho)$.
It should be noted that for $\rho_0 > 1/2$, the dark and bright solitons switch their roles. In view of the 
particle-hole duality, we will present our results for $\rho_0 < 1/2$ where bright solitons have the persistent
character noted above. Note that for $v > c_s$, $\Gamma$ becomes purely imaginary and equation (6) then describes an
infinite periodic train of soliton.

\begin{figure}[htbp]
\includegraphics[height =1.0\linewidth]{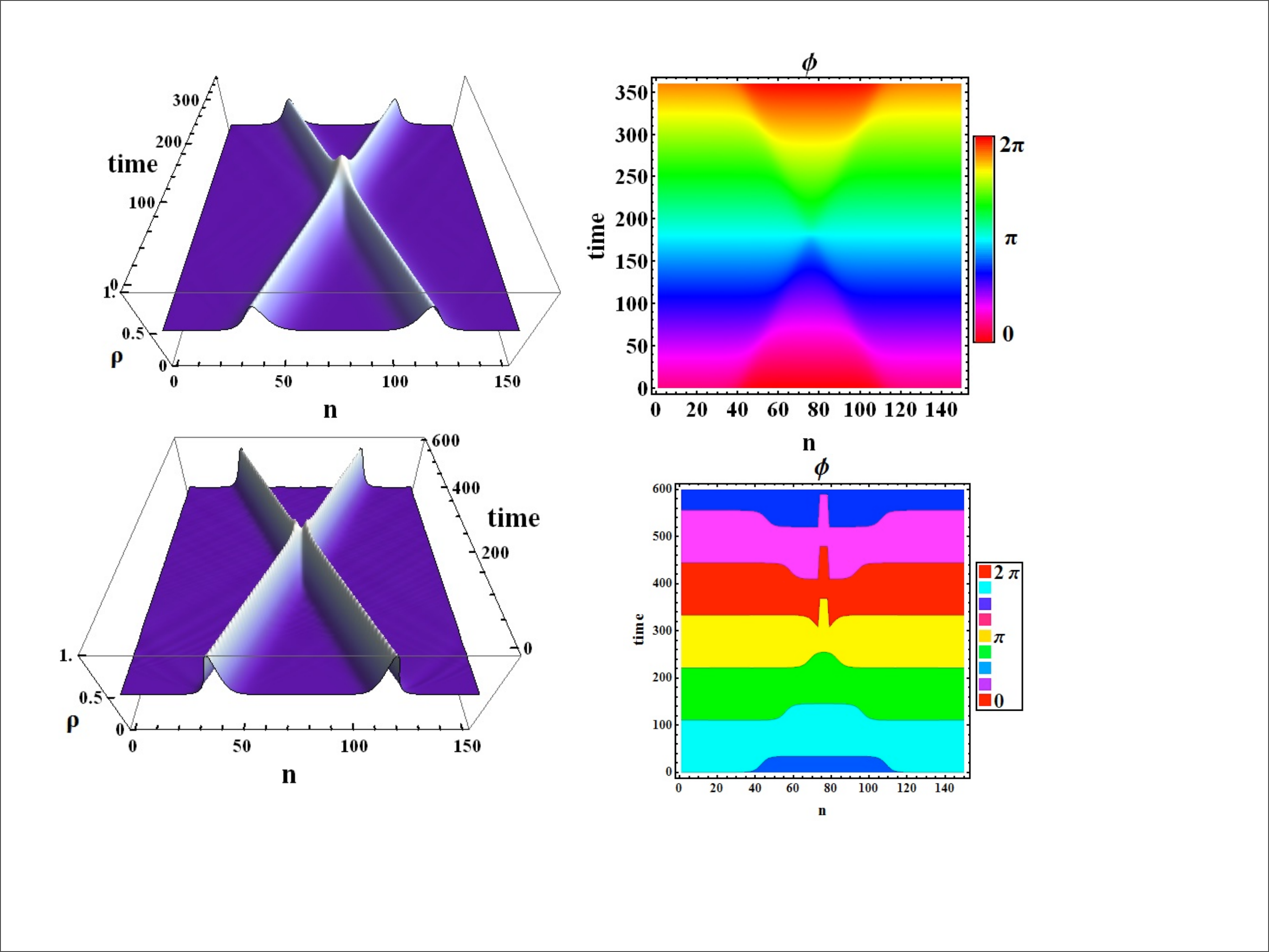}
\leavevmode \caption{(color online) Soliton 3D density profile (left) and its phase (right) projected into 2D illustrating collisions where solitons
pass through ( top, $\bar{v}=.85$) or repel (bottom, $\bar{v}=.5$) each other. Here $V=.9$ and $\rho_0=.45$. As seen from the figure 
the exact soliton collision time can be precisely determined by monitoring the phase: In T-type, it becomes
uniform throughout the lattice at the instant of maximal overlap while in R-type, the collision is signaled by a phase jump 
of $\pi$, a {\it phase wall}, see also Figure 3B for another view of this phase wall.}
\label{TRclass}
\end{figure}


\begin{figure}[htbp]
\includegraphics[width =1.0\linewidth,height=1.3\linewidth]{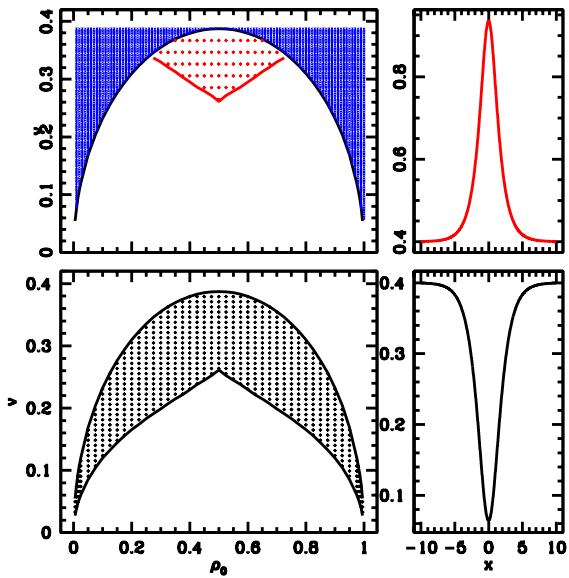}
\leavevmode \caption{(color online) Soliton-soliton collision phase diagrams, top and bottom left hand figures.  The dotted regions indicate parameter regimes where solitons engage in T class collisions (see Fig. 1), and
appear to pass through each other; the white regions below these are those of the R class collisions. as in Fig. 2. The parabolic black line shows the sound velocity $c_s$ as a function of $\rho$.
The blue dots show the supersonic region, applicable only for bright soliton trains.
Top and the bottom figures on the right respectively correspond to
typical bright and the dark individual solitons both shown for $\rho=.45$ and $\bar{v}=.5$, respectively.}
\label{collisionD}
\end{figure}

We now report results of a detailed study of collisions of identical solitary waves, and mirror image bright-dark pairs, moving with opposite velocities
on a lattice, based on integration of
equations (\ref{canonical}) in time, with equation (\ref{feqn}) used to construct initial conditions.
We first discuss intra species collisions of two solitary waves. Figures (\ref{TRclass}) present examples of the two distinctive types of such collisions, referred to here as the T, complete 'Transmission,'  and R, partial 'Reflection,' classes respectively. These collisions are both characterized
by {\it stationary density } , when at some time $t$,  
$\frac{\partial{\rho_j}}{\partial{t}}=0$ for all $j$. 
In the T-class, where solitary waves appear to pass through each other,
the phase $\phi$ becomes uniform across the lattice, 
( from eq. (\ref{canonical}).
In the R-class, where solitary waves appear to repel each other,
$\phi(j)$ is uniform across the lattice except at two singular lattice sites, where there is an anti-node in the density
and a corresponding phase jump of $\pi$ , as encountered in wave reflection phenomena.
The distance between the nodes can be tuned by changing the interaction $V$.
Both classes display a stationary density profile at the collision center, and for the R-class  the particle or hole densities
always attain the maximum density (the anti-node) permitted by the hard core constraint.
In all of our investigations, the solitary waves were found to be solitons: they emerge
intact after collisions. The numerical results presented fully validate the continuum limit solutions equation (\ref{feqn}). 
The equivalent collisions for dark-dark pairs are not shown, but mirror those of Figs.(\ref{TRclass}) , as density notch collisions where the anti-node maxima become actual nodes (zeros) in the superfluid density, and have corresponding phase jumps of $\pi$, as would be the case for an ordinary node in a Schrodinger wave-function. 

We summarize the characteristics of bright and dark pair soliton collisions in a {\it collision phase diagram} 
as shown in Figure (\ref{collisionD}).
The bright solitons repel each other ( R class) provided their their velocity is below a threshold value.
This is rooted in the fact that at low superfluid density, where either the particle or the hole population is large,
the hard core constraint suppresses T class collision.
For dark solitons, on the other hand, there exists a critical velocity for T class collisions for all
values of the background density.
In other words, both bright and dark solitons
encode the hard core constraint of the underlying Hamiltonian.
The supersonic solitons are always in the T-class,
illustrating a remarkable {\it adaptability} of the hard core constraint.

In the case of half-filling, bright and dark solitons are mirror images. Then,
interspecies collisions provide 
a conjugate scenario where the collision center
is characterized by a stationary phase,
$\frac{\partial{\phi_j}}{\partial{t}}=0$ for all $j$ 
,rather than the stationary density of intraspecies collisions.
The colliding solitons temporarily exactly annihilate each other in terms of density profile, but with a strongly variable phase profile, which allows them to re-emerge as counter propagating density defects of opposite sign. Reminiscent of the soliton-antisoliton pair of the sine-Gordon system\cite{book}, the HCB soliton carries a topological charge that is encoded in the the phase jump. 

\begin{figure}[htbp]
\includegraphics[width =1.1\linewidth,height=1.1\linewidth]{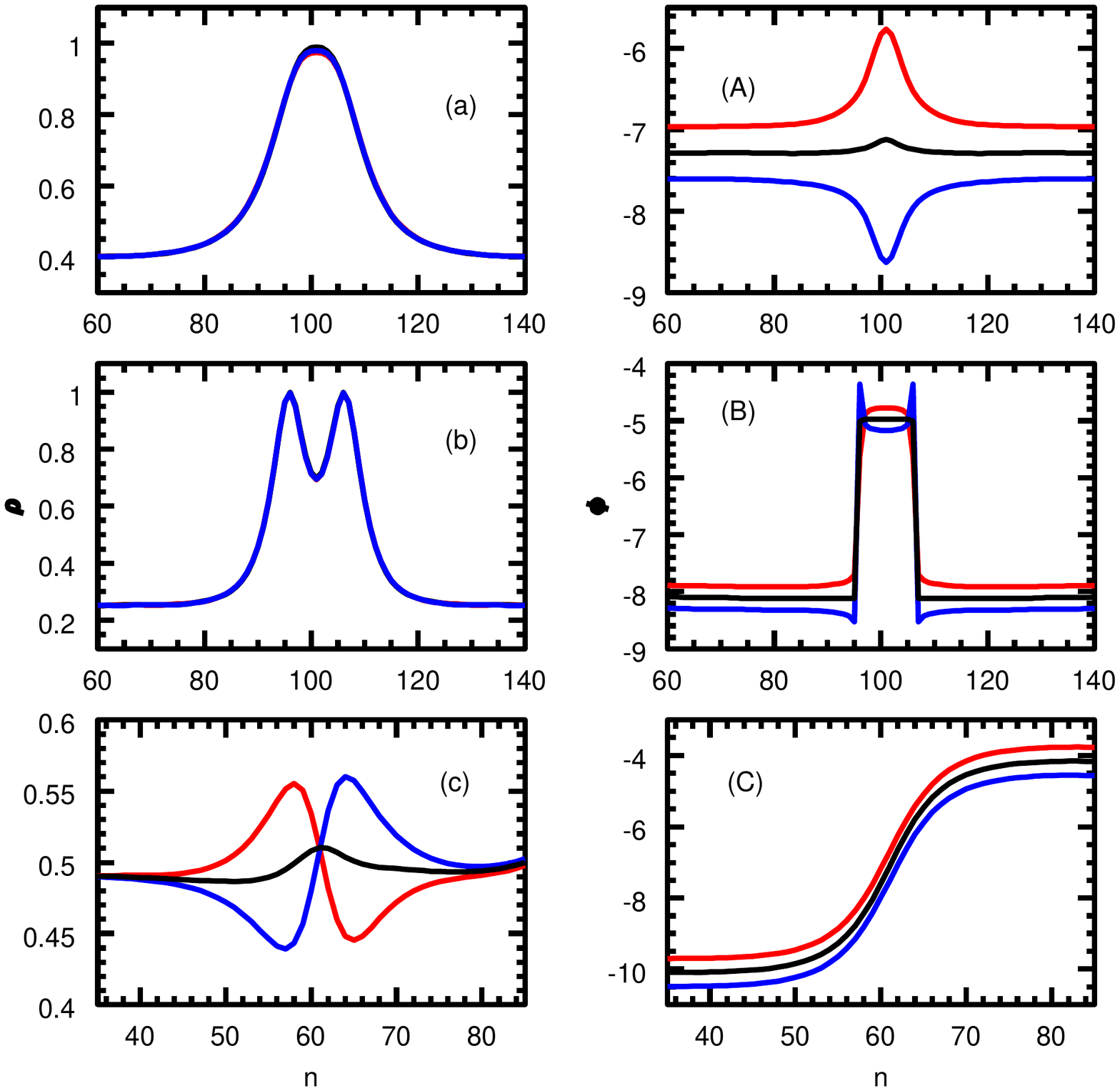}
\leavevmode \caption{(color online)  Soliton density profiles (left) and phases (right) for T-type (a,A) and R-type (b,B) 
bright-bright soliton collisions, and a bright-dark (c,C) interspecies collision.  Shown are denisties and phases before (red), at (black), and after (blue)
the collision time, the blue and red curves being at times  $\pm 4$
 time units with respect to the exact collision time, as {\it defined} by the stationarity of the phase density or phase respectively, for the intra and inter species collisions.}
\label{inter}
\end{figure}

As Fig. (\ref{inter}) shows, during a bright-dark soliton collision of mirror image pairs
there is a smooth phase jump of approximately $2\pi$ across the lattice even at the moment of annihilation.
Thus phase imprinting of a two-$\pi$ phase jump of uniform gas should produce a soliton-antisoliton pair.

\begin{figure}[htbp]
\includegraphics[width =1\linewidth,height=1\linewidth]{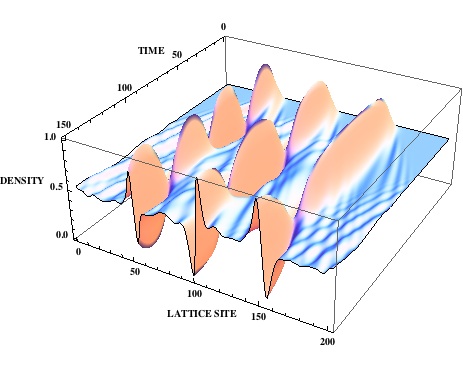}\\
\includegraphics[width =0.8\linewidth,height=0.8\linewidth]{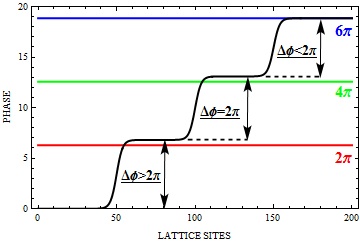}
\leavevmode \caption{(color online) On the left, three breathers, of strikingly different frequencies, 
created using constant density and phase $\phi= n\pi\tanh((x-x_i)/\Gamma)$, with $n=2$, 
for constant $\Gamma =3$ lattice units, and three different values of $x_i$; the actual phase imprinting profiles are show in the right hand inset:  the possibility of startlingly sensitive phase control of these non-linear excitations is evident.}
\label{br2}
\end{figure}

The uniform density and smooth phase profile scenario as illustrated in (\ref{inter}) opens a possibility
to create {\it breathers}\cite{breather}: spatially localized modes that oscillate in time and which may be thought of a bound
state of soliton and antisoliton.
We have found a specific phase imprinting condition that creates such breathers.
An initial condition of 
uniform density at half filling of the lattice and phase profiles of the form $ n\pi\tanh((x-x_i)/\Gamma)$ , 
where the integer $n/2$ determines the number of breathers
in the multi-soliton profile near $x_i$.  See figure (\ref{br2}). These breathers, 
can be dissociated into bright-dark soliton pairs by tuning the strength of the attractive interaction V in the Hamiltonian of Eqn 1.
Small perturbations on the phase profile as shown in figure  (\ref{br2}) lead to
a spectrum of stable oscillatory modes of sharply differing frequencies, each resembling a bound state of bright-dark soliton bound pair.

The HCB constraint prohibits particles moving in one-dimension from passing through each other.
Oddly, though, in the HCB systems treated here, we find
solitons and soliton trains that easily pass through each other in low and high (supersonic) speed collisions.
At low lattice fillings, the system also exhibits solitons that encode
the fermionization characteristics of HCB, and that prohibit transmissive collision.
Unusually, unlike the GPE solitons of gaseous BECs, both bright and dark solitons can be found in the same system, 
for a fixed set of parameters in the lattice Hamiltonian.
Non-GPE-type solitary waves have been the subject of various recent studies.\cite{Kolo,Lew}. The present studies
add a new class of solitons and soliton trains relevant to strongly interacting BEC.
The existence of solitons in mean-field equations for HCB,
perhaps has its roots in the integrability of the underlying XXZ spin Hamiltonian
in a magnetic field.\cite{BA} 
Our numerical studies suggest the existence of a variety of novel multi-solitons that include double-node (or double antinode) solitons that repel, and of a broad class of breathers whose frequencies and integrations may be controlled in the laboratory by quantum phase engineering.

We would like to thank Erhai Zhao for various comments and suggestions. The research of IIS is supported by
ONR grant N00014-09-1-1025A and grant 70NANB7H6138 Am 001 by NIST and that of WPR by NSF PHYS 07-03278.

\end{document}